\documentclass[twocolumn,english]{revtex4-1}
\usepackage[T1]{fontenc}
\usepackage[latin9]{inputenc}
\usepackage{amstext}
\usepackage{graphicx}
\usepackage{esint}
\usepackage{babel}
\begin{document}

\title{Shape minimization problems in liquid crystals}

\author{Andrew DeBenedictis}

\affiliation{Department of Physics and Astronomy, Tufts University, 574 Boston
Avenue, Medford, Massachusetts 02155, USA}

\author{Timothy J. Atherton}

\affiliation{Department of Physics and Astronomy, Tufts University, 574 Boston
Avenue, Medford, Massachusetts 02155, USA}
\begin{abstract}
We consider a class of liquid crystal free-boundary problems for which
both the equilibrium shape and internal configuration of a system
must simultaneously be determined, for example in films with air-
or fluid-liquid crystal interfaces and elastomers. We develop a finite
element algorithm to solve such problems with dynamic mesh control,
achieved by supplementing the free energy with an auxiliary functional
that promotes mesh quality and is minimized in the null space of the
energy. We apply this algorithm to a flexible capacitor, as well as
to determine the shape of liquid crystal tactoids as a function of
the surface tension and elastic constants. These are compared with
theoretical predictions and experimental observations of tactoids
from the literature. 
\end{abstract}
\maketitle

\section{Introduction}

Surface effects are of fundamental importance in liquid crystal physics\cite{barbero2005adsorption,rasing2013surfaces}.
Chemical and topographical treatments induce a preferred molecular
orientation that, due to elasticity, tends to align the material in
the bulk. Patterned surfaces produce a wealth of phenomena, including
multistable alignment\cite{Kim2002,Atherton2012} and arbitrary control
of the alignment orientation and associated anchoring energy\cite{Kondrat2005,Anquetil-Deck2013}.
Highly confined liquid crystals, in channels, emulsion droplets, or
polymer dispersed systems are also strongly influenced by the interaction
of the LC material with the adjacent medium. As liquid crystals are
deployed beyond display applications\cite{lagerwall2012new}, the
exploration of their behavior in complex geometries has become increasingly
important. Almost ubiquitously, the shape of their enclosure is regarded
as a rigid body and the equilibrium configuration of the liquid crystal
found by minimizing an appropriate energy functional, a challenging
enough task given the multiple competing physical effects\textemdash elasticity,
surfaces, applied and internal fields\textemdash involved. If the
external environment is not rigid, the minimization problem also requires
solving for the shape of the body. It is this latter category that
shall be discussed in this paper, which we refer to as shape-order
problems since they require simultaneous solution of both shape and
ordering. 

One situation where the variable shape problem may occur is when droplets
or shells of liquid crystalline phase are dispersed in an isotropic
fluid. For thermotropic liquid crystals, the nematic-isotropic interfacial
energy is much larger than the elastic energy, and such droplets remain
approximately spherical. Topology restricts the number and type of
defects on the surface\cite{MacKintoshLubensky1991,fernandez2007novel},
which can be used to template other objects \cite{PhysRevLett.111.227801}.
For lyotropic materials, however, the effective energy of the phase
boundary is weak and highly deformed droplets, known as \emph{tactoids,
}are observed. Observed in dilute vanadium pentoxide by Zocher~\cite{Zocher1925,Zocher1929}
in 1925 and assemblies of viral particles Bernal and Fankuchen~\cite{Bernal1941}
in 1941, there has been a resurgence of interest in the last couple
of decades and a rich variety of morphologies observed. In an applied
electric field, paramagnetic tactoids are deformed strongly as a flow
field develops internally~\cite{Sonin1998}. While most tactoids
observed possess a characteristic eye shape, a rare triangular version
has been seen using an actin system with crosslinkers~\cite{Tang2005}.
Non simply-connected shapes are commonly seen, with the isotropic
inclusions themselves forming negative tactoids~\cite{Casagrande1987}
and examples with many inclusions and defects have been seen in chromonic
liquid crystals~\cite{Kim2013}. While for most tactoids the anchoring
is parallel to the interface, elongation has also been seen in small
droplets with perpendicular (homeotropic) anchoring~\cite{Verhoeff2011}.
Tactoid structures have been shown to act as sensors via chirality
amplification~\cite{Peng2015} and can be used to guide motile bacteria~\cite{Mushenheim2014,Zhou2014}.
They are also valuable architectural elements of self assembly, for
example providing nucleation sites for growth of the smectic phase~\cite{Dogic2001}. 

Theoretical treatments of tactoids are more sparse due to the difficulty
of the combined shape-order minimization problem. Chandrasekhar argued
for a taxonomy of four possible shapes~\cite{Chandrasekhar1966}:
elliptical, rectangular, a rounded rectangle and the eye shape observed
in earlier experimental work. Kaznacheev proposed an elastic theory~\cite{Kaznacheev2003}
for the size-dependence of the equilibrium shape, determining that
larger tactoids become more spherical in agreement with experiments~\cite{Kaznacheev2002}.
Prinsen argued that the aspect ratio of the tactoid depends primarily
on the anchoring strength while the configuration of the director
field depends on the ratio of anchoring to elastic constant~\cite{Prinsen2003}.
In a second paper, they showed that differences between elastic constants
can lead to several different stable director configurations and predicted
a phase diagram for the ground state~\cite{Prinsen2004a}. Simulation
based approaches have been used to study tactoids. Bates used a Monte
Carlo approach to show the tendency of droplets to elongate~\cite{Bates2003}.
Recently, a coarse grained simulation~\cite{Vanzo2012} of $50-100nm$
droplets showed that the liquid crystal may spontaneously undergo
a symmetry breaking transition to a chiral configuration in a tactoid,
despite being composed of achiral materials. 

A second class of materials that undergo dramatic shape change are
Liquid Crystal Elastomers (LCEs), soft polymeric materials consisting
of a cross-linked network and rod-like mesogens that spontaneously
align. These materials combine the properties of rubbery materials
and liquid crystals~\cite{Warnerbook}. Their chief distinctive feature
is the coupling between strain and orientation order: stretching a
LCE material can rotate the mesogens and vice versa~\cite{DeGennes,kundler_finkelmann,Finkelmann2001,Conti2002,Mbanga_Selinger_pre}.
They may be induced to undergo large deformations by external stimuli~\cite{Finkelmann2001}
including electric fields~\cite{Chang1997,Burridge2006,Menzel2007,Ye2007},
induced swelling by absorption of solvent molecules \cite{Urayama2003,Urayama2006}
and light~\cite{Yu2003,azo_Keller_2003,Camacho-Lopez2004,He2007,Sawa2010,azo_review_Barrett}.
These fascinating properties suggest many applications requiring flexibility
or a large response to a modest stimulus, such as soft actuators,
artificial muscles, self-folding structures, smart materials, flexible
displays, and flexible microgenerators~\cite{Ikeda,MASY:MASY200351008,Palffy-Muhoray2009,smart_materials}.
In light of the strong contemporary interest in these materials, some
groups have begun to produce high quality numerical solvers to simulate
their behavior\cite{mbanga2010modeling,ADFM:ADFM201302568}. 

In light of the above applications in liquid crystal physics, as well
as others in soft matter more broadly, there is a need for generic
tools to solve shape-order problems. Excellent tools for shape minimization
problems, such as the \emph{Surface Evolver}\cite{brakke1992surface},
exist but are limited to systems without order and to problems involving
low dimensional manifolds. Unfortunately, these extensions make the
problem much more challenging because the numerical representation\textemdash i.e.
the mesh for a finite element approach\textemdash must be co-evolved
during the minimization along with the boundary. In this paper, we
discuss some initial explorations and propose a regularization scheme
to address these difficulties. The paper is structured as follows:
in section \ref{sec:Model} the numerical challenges in this class
of problem are discussed and a regularization strategy proposed. In
section \ref{sec:Applications} we apply this algorithm to two problems
of increasing difficulty: in subsection \ref{sub:A-flexible-capacitor}
a flexible capacitor for which an analytical solution is available
and in subsection \ref{sub:Tactoids} liquid crystal tactoids. Conclusions
and prospects for future work are presented in section \ref{sec:Conclusion}.

\section{Regularizing Shape Minimization Problems\label{sec:Model}}

As an illustration of the challenges associated with shape minimization
problems, consider perhaps the simplest possible example: minimize
the length of a loop subject to a constant enclosed area of unity.
The problem may be solved in discrete form by defining linear elements
between adjacent points $\mathbf{x}_{i}$ with $i\in[1,N]$ and imposing
a periodic condition $i+N\equiv i$. The program is then,
\begin{eqnarray}
\min & \sum_{i=1}^{N}\left\Vert \mathbf{x}_{i+1}-\mathbf{x}_{i}\right\Vert \nonumber \\
\textrm{s.t.} & \ \frac{1}{2}\sum_{i=1}^{N}\left\Vert \mathbf{x}_{i}\times(\mathbf{x}_{i+1}-\mathbf{x}_{i})\right\Vert  & =1.\label{eq:program}
\end{eqnarray}
By differentiating the target and constraint functionals, we obtain
generalized forces $\mathbf{F}_{i}^{L}$ and $\mathbf{F}_{i}^{A}$
representing the generalized force on vertex $i$ exerted by the length
functional and area constraint respectively. From an initial configuration
$\mathbf{x}^{0}$, the program (\ref{eq:program}) is solved by taking
gradient descent steps,
\begin{equation}
\mathbf{x}^{n+1}=\mathbf{x}^{n}+\lambda\left(\mathbf{F}^{L}-\frac{\mathbf{F}^{L}\cdot\mathbf{F}^{A}}{\mathbf{F}^{A}\cdot\mathbf{F}^{A}}\mathbf{F}^{A}\right),\label{eq:gradientdescent}
\end{equation}
where the component of $\mathbf{F}^{L}$ that tends to change the
area is subtracted off. The stepsize $\lambda$ is determined by a
line search, i.e. performing a 1D minimization of the energy functional
with respect to $\lambda$. After each step, the solution is rescaled
to have unit area. Although simple, this algorithm is at the core
of programs like \emph{Surface Evolver }whose complexity arises from
the wide variety of energy functionals available. To obtain a solution,
the user performs a succession of relaxation steps from an initial
guess to find a solution and is able to interactively refine the mesh.

\begin{figure}
\includegraphics[width=1\columnwidth]{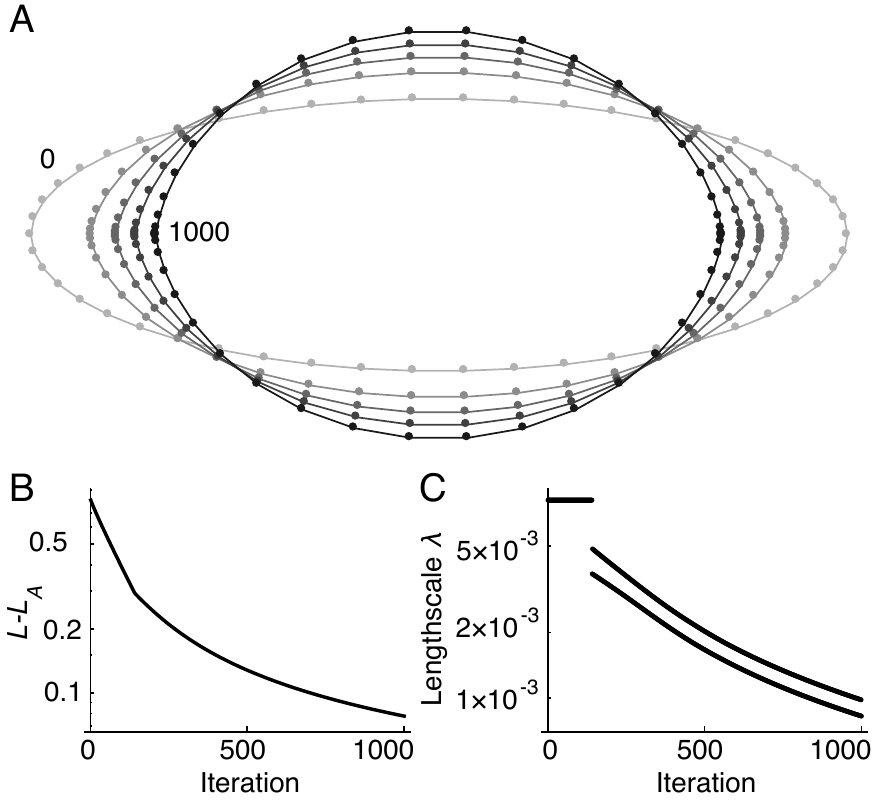}

\caption{\label{fig:Noreg}Minimization of the length of a loop at constant
area. \textbf{A} Intermediate steps at iterations $0$, $100$, $200$,
$400$ and $1000$. \textbf{B }Length of the line minus the analytical
perimeter $L_{A}$ as a function of iteration. \textbf{C} Stepsize
determined from linesearches as a function of iteration. }
\end{figure}

If the initial guess is far from the solution, however, problems can
occur. In fig. \ref{fig:Noreg} we display a minimization starting
from an initial elliptical loop of aspect ratio $3$ and unit area.
The solution converges toward the correct circular solution, but very
slowly. In fig. \ref{fig:Noreg}A, note that mesh points that started
at the ends of the ellipse become bunched together. Hence, even after
$1000$ iterations, only a relatively poor approximant to the circle
has been found. In fig. \ref{fig:Noreg}B the quantity $L-L_{A}$
is reported on log-linear scales as a function of iteration number,
where $L$ is the length of the loop and $L_{A}=2\sqrt{\pi}$ is the
analytical perimeter of a circle with unit area. For a good initial
guess, this quantity should converge exponentially, i.e. $L-L_{A}\propto\exp\left(-kn\right)$
with some rate $k$ and hence would appear on this plot as a straight
line; the line observed shows that the convergence is subexponential.
The reason for this is apparent from a plot of the stepsize $\lambda$
as a function of iteration, shown in fig. \ref{fig:Noreg}C, that
shows that the algorithm is forced to take smaller and smaller steps
as the points become bunched together. 

Bunching arises because the solution of the program (\ref{eq:program})
possesses a null space with respect to the position of the vertices.
Suppose all vertices other than $\mathbf{x}_{i}$ are fixed, but the
$i$th vertex is perturbed $\mathbf{x}_{i}\to\mathbf{x}_{i}+\mathbf{\delta}\mathbf{m}$
where $\mathbf{m}$ is a unit vector. Expanding the length of a loop
as a series in $\delta$ we obtain, 
\begin{equation}
L=L_{0}+\delta\mathbf{m}\cdot\left(\frac{\mathbf{x}_{i+1}-\mathbf{x}_{i}}{\left|\mathbf{x}_{i+1}-\mathbf{x}_{i}\right|}-\frac{\mathbf{x}_{i}-\mathbf{x}_{i-1}}{\left|\mathbf{x}_{i}-\mathbf{x}_{i-1}\right|}\right)+\cdots.\label{eq:nullspace}
\end{equation}
The change in length can be cancelled to linear order if the direction
of the perturbation is chosen to be perpendicular to the second factor
in the dot product. The factor in question is, by definition, the
generalized force acting on the vertex $i$. At equilibrium, this
must balance the force due to the area constraint, which always points
along the normal. Hence, for the solution of (\ref{eq:program}),
the nullspace lies for each vertex locally along the tangent line. 

The existence of the nullspace reveals that the problem is underconstrained,
and this characteristic is generic for shape evolution problems. For
instance, if the interior of the loop were to be meshed as well, the
position of these additional mesh points does not enter the program
(\ref{eq:program}), and so they would not be moved. Moreover, it
is an issue fundamentally associated with the discrete system as the
continuous functional possesses a unique solution. 

\begin{figure}
\includegraphics[width=1\columnwidth]{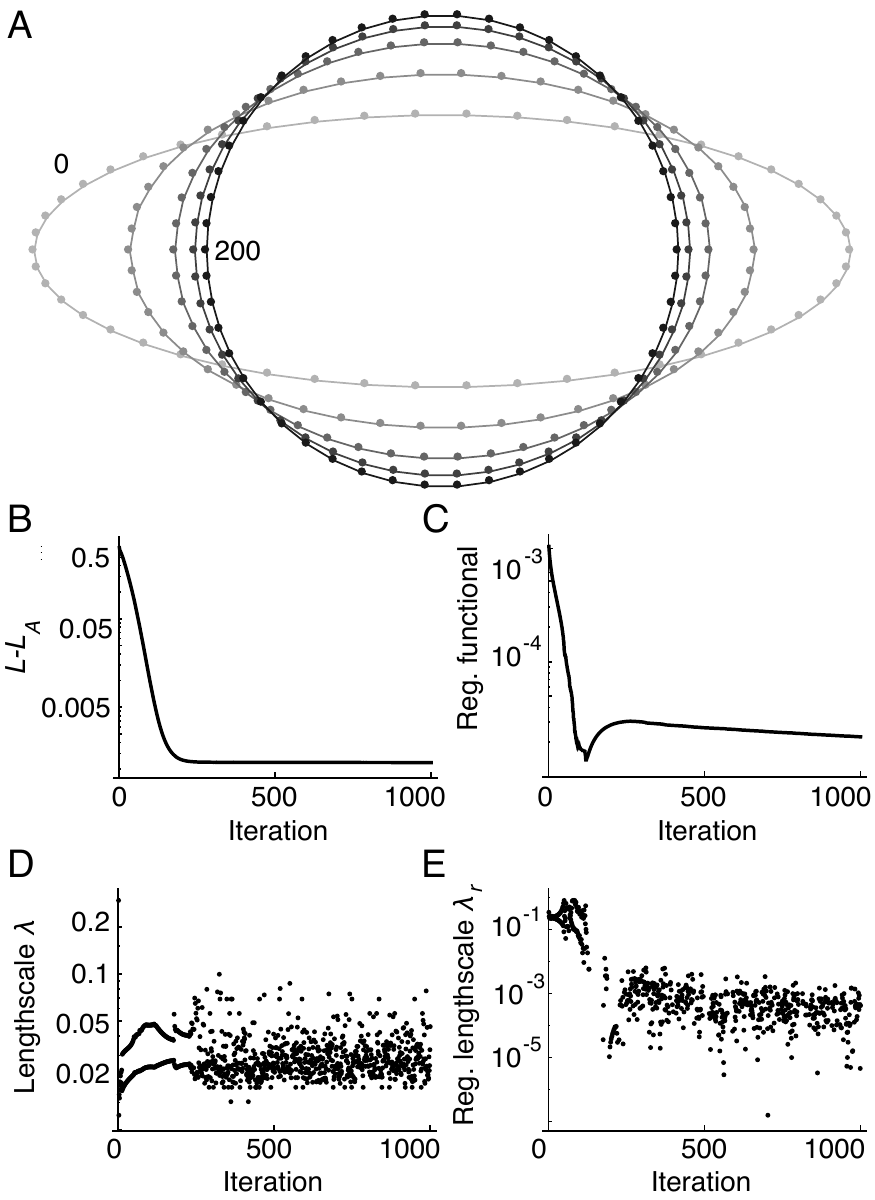}

\caption{\label{fig:Reg}Minimization of the length of a loop with regularization.
\textbf{A }Intermediate steps at iterations $0$, $40$, $80$, $120$
and $200$. \textbf{B }Length of the line minus the analytical perimeter
$L_{A}$ as a function of iteration. \textbf{C }Value of the regularization
functional as a function of iteration. \textbf{D} Stepsize $\lambda$
and \textbf{E} Regularization stepsize $\lambda_{r}$ as determined
from linesearches by iteration. }
\end{figure}

To solve this problem, \emph{Surface Evolver} permits the user to
merge nearby mesh points as they become bunched, and to create new
mesh points by splitting long elements. This is effective, but means
that the number of mesh points cannot be controlled as easily and
requires discrete changes to the mesh. Here, we propose a different
solution that allows continuous regularization of the mesh. The idea
is to supplement the minimization problem with an auxiliary functional
that captures our intuition of mesh quality. For the present problem,
a suitable functional is,
\begin{equation}
E_{R}=\frac{1}{2}\sum_{i}\left[\left(\left|\mathbf{x}_{i}-\mathbf{x}_{i-1}\right|-\bar{L_{i}}\right)^{2}+\left(\left|\mathbf{x}_{i+1}-\mathbf{x}_{i}\right|-\bar{L_{i}}\right)^{2}\right]\label{eq:regularization}
\end{equation}
where $\bar{L_{i}}=\frac{1}{2}\left(\left|\mathbf{x}_{i}-\mathbf{x}_{i-1}\right|+\left|\mathbf{x}_{i+1}-\mathbf{x}_{i}\right|\right)$,
i.e. the mean length of the two linear elements at point $i$. This
functional favors equally spaced mesh points. By differentiating the
regularization functional, a corresponding generalized force $\mathbf{F}^{R}$
may be obtained. 

It is essential that the regularization functional does not interfere
with the solution of the program (\ref{eq:program}), so the regularization
functional is minimized separately in the nullspace of the target
functional, i.e ensuring that $\mathbf{F}^{L}\cdot\mathbf{F}^{R}=\mathbf{F}^{A}\cdot\mathbf{F}^{R}=0$.
Following each gradient descent step eq. (\ref{eq:gradientdescent}),
a regularization step is then taken,

\begin{equation}
\mathbf{x}^{n+1}=\mathbf{x}^{n}+\lambda_{r}\left(\mathbf{F}^{R}-\frac{\mathbf{F}^{R}\cdot\mathbf{F}^{L}}{\mathbf{F}^{L}\cdot\mathbf{F}^{L}}\mathbf{F}^{L}-\frac{\mathbf{F}^{R}\cdot\mathbf{F}^{A}}{\mathbf{F}^{A}\cdot\mathbf{F}^{A}}\mathbf{F}^{A}\right),\label{eq:gradientdescent-1}
\end{equation}
where the regularization stepsize $\lambda_{r}$ is found by performing
a linesearch on the regularization functional (\ref{eq:regularization}). 

Execution of this algorithm on the same elliptical initial configuration
as above yields a plot displayed in figure \ref{fig:Reg}A that has
converged onto the correct circular solution with the mesh points
are approximately evenly spaced as desired. Inspection of the length
as a function of iteration $L-L_{A}$ in figure \ref{fig:Reg}B shows
that the algorithm converges exponentially, within around $200$ iterations,
even from this poor initial guess. While the new algorithm involves
approximately twice the work per iteration, i.e. an additional force
evaluation and line search, this is clearly a considerable improvement
over the naive algorithm. The value of the $E_{R}$ is shown in figure
\ref{fig:Reg}C indicating that the solution is not fully converged
with respect to the regularization functional. While the stepsizes
$\lambda$ and $\lambda_{R}$\textemdash displayed in figure \ref{fig:Reg}D
and \ref{fig:Reg}E respectively\textemdash are very different in
value and rather noisy, it is apparent that they remain approximately
constant with iteration, explaining the fast convergence. 

The strategy developed for the very simple problem considered in the
present section is readily usable to any combination of target functionals,
constraints and regularization functionals. In particular, it is also
applicable to problems where the minimization must be performed not
only on the shape of the domain but on field quantities defined upon
it; it is to such a problem that we turn in the following section.

\section{Applications\label{sec:Applications}}

\subsection{A flexible capacitor\label{sub:A-flexible-capacitor}}

\begin{figure}
\includegraphics[width=1\columnwidth]{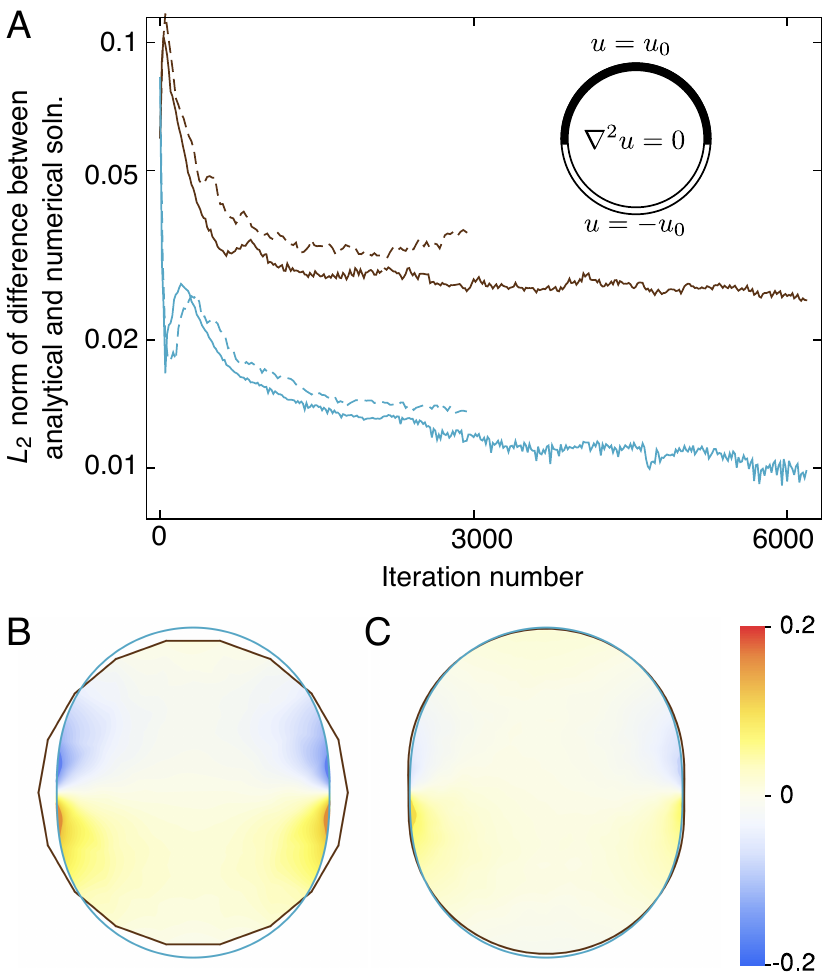}

\caption{\label{fig:Conformal}Convergence to the conformal mapping solution
for a flexible capacitor with $\Gamma_{\sigma}=2$ and $\Gamma_{W}=1$.
\textbf{A} Norm of the difference between simulation and conformal
mapping results for the field (brown) and shape (cyan) for simulations
with (solid) and without (dashed) the regularization routine as the
simulation reaches convergence. Comparison plots of the \textbf{B}
initial and \textbf{C} final states of the simulation. The outer borders
of the CM (cyan) and simulation (brown) solutions are shown, and points
interior to both shapes are shaded by the difference between the field
solutions at those points.}
\end{figure}

We now examine a more complex problem, a cylindrical capacitor made
from flexible electrodes shown schematically in the inset of fig.
\ref{fig:Conformal}A. Normal capacitors with rigid plates held at
fixed potential difference experience a force that tends to separate
the plates. Here, the capacitor is initially circular in the absence
of an electric field and tends to elongate in the vertical direction
when a potential difference is applied to the terminals. The device
is filled with an incompressible fluid dielectric and so maintains
a constant area. 

The energy functional contains three terms, a line tension that minimizes
droplet circumference, the electrostatic energy resulting from the
electric field, and an anchoring energy used to enforce a boundary
value $u=u_{0}$ without discontinuities on the electric potential,
\begin{equation}
E=\sigma\int_{\partial D}dl+\Lambda\epsilon\int_{D}(\nabla u)^{2}dA+W\int_{\partial D}(u-u_{0})^{2}dl.\label{eq:energy-1}
\end{equation}
This is to be minimized on a 2D domain $D$ with respect to the electric
potential $u$ and the shape of the boundary $\partial D$, subject
to fixed area,
\begin{equation}
\int dA=\pi.\label{eq:areaconstraint}
\end{equation}
The parameters $\sigma$, $W$ and $\epsilon$ represent the surface
tension, anchoring and electrostatic energies respectively; $\Lambda$
is a length of the order of the size of the droplet. Note that this
problem is easily transformed into a liquid crystal problem by replacing
the electric potential with the director angle $\theta$, modifying
the elastic term to include variable elastic constants and replacing
$u_{0}$ with a spatially varying easy axis. 

The energy (\ref{eq:energy-1}) can be made dimensionless by dividing
through by $\Lambda\epsilon u_{0}^{2}$, and making a change of variables
$x\rightarrow\Lambda x'$,
\begin{equation}
\frac{E}{\Lambda\epsilon u_{0}^{2}}=\frac{\sigma}{\epsilon u_{0}^{2}}\int_{\partial D}dl'+\int_{D}(\nabla'\overline{u})^{2}dA'+\frac{W}{\epsilon}\int_{\partial D}(\overline{u}-1)^{2}dl',\label{eq:energy-noDim}
\end{equation}
where $\overline{u}=u/u_{0}$ and the primed variables correspond
to the coordinate system scaled by $\Lambda$. There are two dimensionless
parameters, $\Gamma_{\sigma}=\frac{\sigma}{\epsilon u_{0}^{2}}$,
which represents the ratio of line tension to applied field, and $\Gamma_{W}=\frac{W}{\epsilon}$,
which represents the relative importance of anchoring and electric
forces. 

The problem is discretized using a conventional finite element approach:
the domain is represented by an initially circular mesh with triangular
elements. The potential $u$ is represented on the vertices and a
linear interpolation is used on the interior of the triangles, hence
$\nabla u$ is constant in each triangular element. Using this discretization,
an estimate of the functional (\ref{eq:energy-noDim}) can be calculated
for a given configuration. Formulae for these expressions are available
in standard finite element textbooks; we do not supply them here.
From these, it is possible to obtain the generalized force $\mathbf{F}$
acting on each vertex of the mesh, and also a scalar generalized force
$F$ acting on the field. Boundary vertices experience a contribution
to the force from all terms in (\ref{eq:energy-noDim}); the interior
vertices only experience forces due to the electrostatic term. Because
of this, it is clear that the problem is underconstrained with respect
to the interior vertices. 

To remedy this, two auxiliary functionals are introduced to regularize
the problem. The first,
\begin{equation}
E_{R,energy}=\sum_{i}^{N_{f}}\left[\int_{D_{i}}(\nabla u)^{2}dA-\frac{\int_{D}(\nabla u)^{2}dA}{N_{f}}\right]^{2},\label{eq:reg1}
\end{equation}
prefers to maintain equal energy density among all triangles; the
second,
\begin{equation}
E_{R,angle}=\sum_{i}^{N_{f}}\left[\frac{1}{\pi}\sum_{j=1}^{3}(\psi_{i,j}-\pi/3)^{2}\right],\label{eq:reg2}
\end{equation}
favors triangles with equal angles. In (\ref{eq:reg1}) and (\ref{eq:reg2}),
$\psi_{i,j}$ is the interior angle of face $i$ at vertex $j$ and
$N_{f}$ is the total number of triangles composing the mesh. Using
the algorithm described in the previous section, gradient descent
is performed by successive line searches on the target and auxiliary
functionals. 

To validate our algorithm, we take advantage the fact that an approximate
analytical solution is available from conformal mapping, at least
in the limit of small deformations. We use a one-parameter area conserving
conformal map $z(w)$ of the form,
\begin{equation}
z=w\sqrt{1-3\alpha^{2}}+\alpha w^{3}.\label{eq:Ourmap-1}
\end{equation}
to map the undeformed capacitor, i.e. the unit disk, parametrized
by complex coordinates $w$, to a target domain with coordinates $z$.
The electric potential is also represented in the source domain with
a complex function $u(w)$,
\begin{equation}
u(w)=\Im\left(\sum_{n=1}^{\infty}c_{n}w^{n}\right),\label{eq:usoln-2}
\end{equation}
with coefficients $c_{n}$. By translating the energy functional (\ref{eq:energy-noDim})
into complex form, inserting the map eq. (\ref{eq:Ourmap-1}) and
potential eq. (\ref{eq:usoln-2}), and performing the integrations,
we obtain an expression for the energy that is quadratic in $\alpha$
and $c_{n}$ and readily minimized numerically. Full details of the
derivation are provided in the appendix. 

\begin{figure}
\includegraphics[width=1\columnwidth]{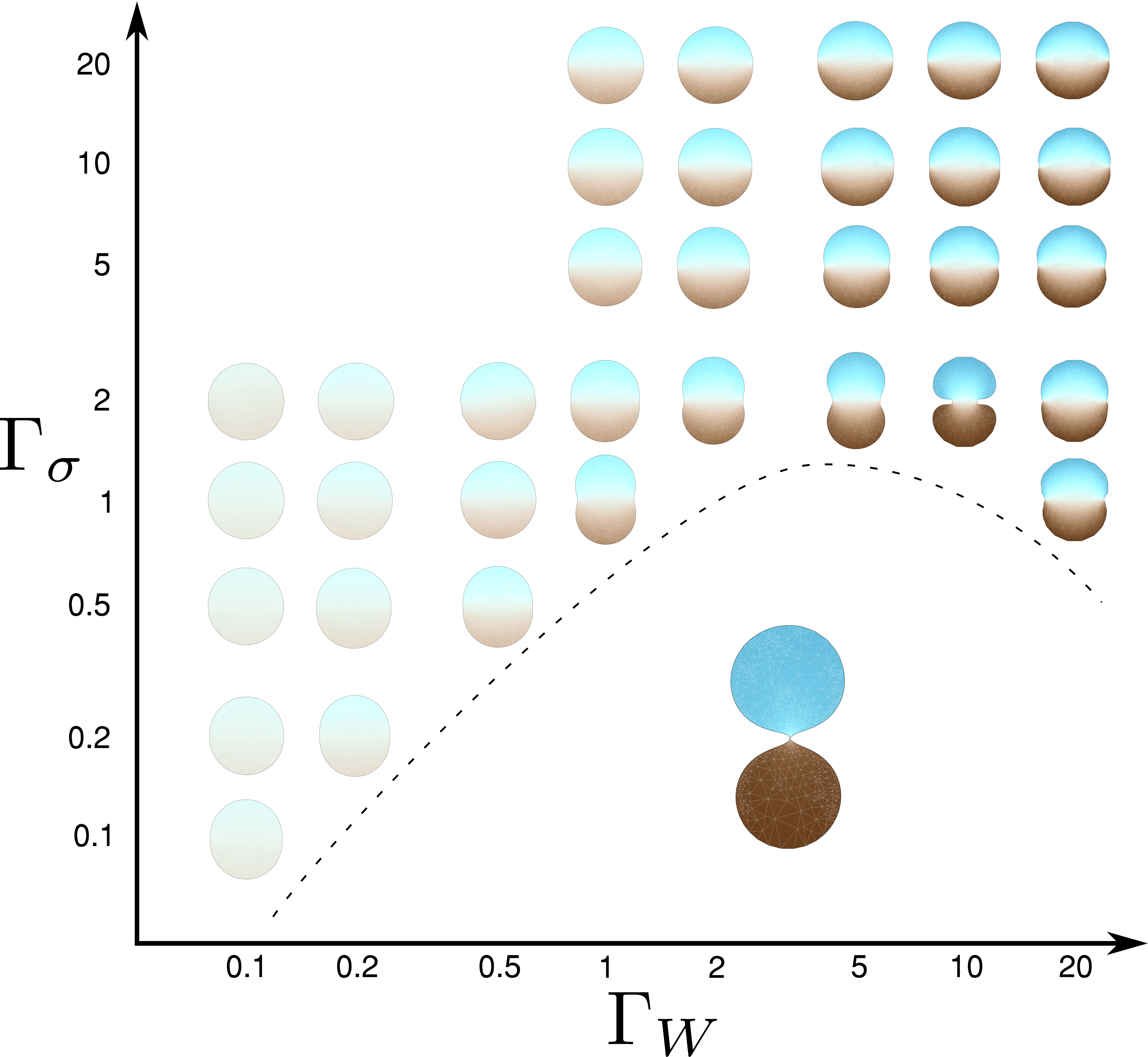}

\caption{\label{fig:FlexibleCapa}Shape of a flexible capacitor with applied
electric field. Deep cyan (brown) regions indicate a field value of
$+u_{0}$ ($-u_{0}$). All cases were started with a circular shape
and a field that changes linearly along the vertical. Solutions below
the dashed line converge on a ``pinched'' configuration that prefers
to separate. }
\end{figure}

For a test case with parameters $\Gamma_{\sigma}=2$ and $\Gamma_{W}=1$,
we show in fig. \ref{fig:Conformal}A a comparison between the conformal
mapping solution and the finite element solution as a function of
iteration for the $L_{2}$ norm of the difference in the field, 
\[
\left\Vert u_{c}-u_{fe}\right\Vert _{2}=\int|u_{c}-u_{fe}|^{2}dA,
\]
and the $L_{2}$ norm of the difference in shape, i.e. the area between
the finite element boundary and the conformal mapping boundary. Both
quantities converge to a fixed nonzero value because the conformal
mapping solution is itself approximate. Additionally, we show these
quantities with and without the regularization scheme active. Note
that without regularization, the method fails to converge due to bunching
of the mesh points. In figure fig. \ref{fig:Conformal}B and fig.
\ref{fig:Conformal}C, a local comparison between the conformal mapping
and finite element solution is shown for the start and end configurations,
showing a good agreement. 

In figure \ref{fig:FlexibleCapa} we display configurations that result
from these minimizations as a function of $\Gamma_{W}$  and $\Gamma_{\sigma}$
including those that exhibit large deformations outside the validity
of the conformal mapping solution. For $\sigma>W$, the capacitors
tend to remain fairly circular; as $\Gamma_{W}$ is increased the
solution conforms more closely to the boundary conditions. Conversely,
for $\sigma<W$ significant deformations of the capacitor are allowed;
the shape elongates vertically at the expense of horizontal invaginations.
Beneath the dashed line in fig. \ref{fig:FlexibleCapa}, the deformation
is so extreme that the invaginations meet; if allowed to, the capacitor
would seperate into two disconnected domains.

\subsection{Tactoids\label{sub:Tactoids}}

\begin{figure}
\includegraphics[width=1\columnwidth]{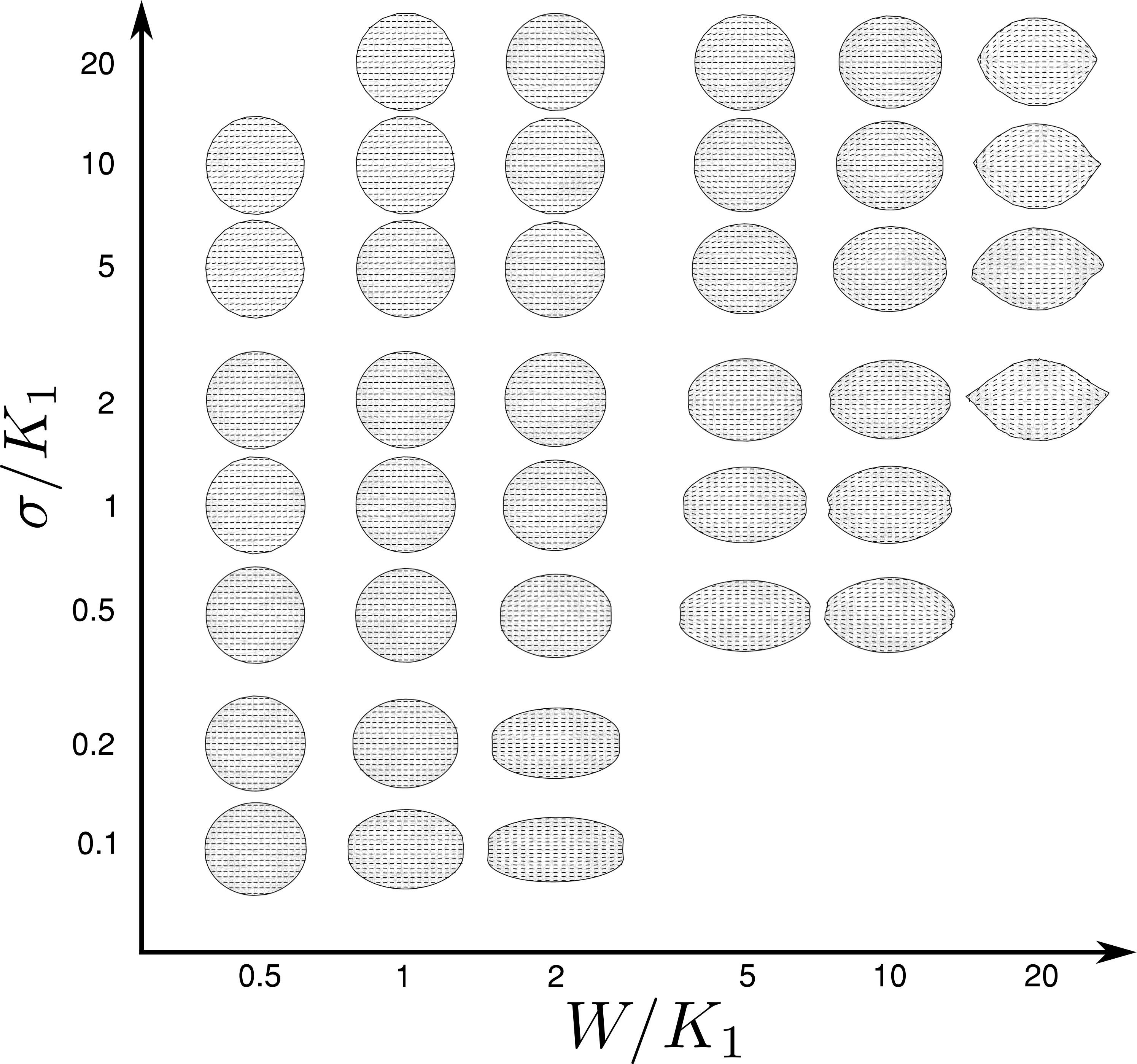}

\caption{\label{fig:Shape-of-a}Shape of a nematic droplet. All cases were
started with a circular shape and a director field with virtualized
defects on the horizontal axis. Depending on the anchoring strength,
these defects may spread out to infinity or may move into the droplet.}
\end{figure}

We now adapt the simulation developed in the previous section to study
nematic tactoids. Parametrizing the director as $\mathbf{n}=(\sin\theta,\cos\theta)$,
the energy functional contains line tension, elastic and anchoring
terms respectively,
\begin{eqnarray*}
E & = & \sigma\int_{\partial D}dl+\Lambda\int_{D}\left[K_{1}\left(\nabla\cdot\mathbf{n}\right)^{2}+K_{3}\left|\nabla\times\mathbf{n}\right|^{2}\right]dA\\
 &  & +W\int_{\partial D}(\theta-\theta_{e})^{2}dl
\end{eqnarray*}
where $K_{1}$ and $K_{3}$ are the splay and bend elastic constants
respectively. This is to be minimized subject to an area constraint
as before, and we divide through by $\Lambda K_{1}$ and make the
change of variables $x\rightarrow\Lambda x'$ in order to non-dimensionalize
the energy.

When $K_{3}=K_{1}$, the elastic energy density reduces to $K_{1}\left(\nabla\theta\right)^{2}$
and the problem is almost exactly the same as that discussed in the
previous section, with one important difference: the easy axis $\theta_{e}$
is not fixed on the boundary, but dynamically determined at each iteration
relative to the local estimate of the tangent plane. Even so, we use
the same regularization functionals and algorithm, only inserting
a modified force functional for the elastic anisotropy. 

In figure \ref{fig:Shape-of-a}, a set of equilibrium shapes is shown
for $K_{3}=K_{1}$ as a function of $\sigma/K_{1}$ and $W/K_{1}$.
In all cases, the minimization was started from a circular initial
configuration, with virtual defects outside the tactoid. Stronger
anchoring causes the nematic to better conform to the boundary condition,
effectively moving the virtual defects closer to the surface of the
droplet. Weaker line tension, on the other hand, permits more elongation
of the droplet. 

\begin{figure}
\includegraphics[width=1\columnwidth]{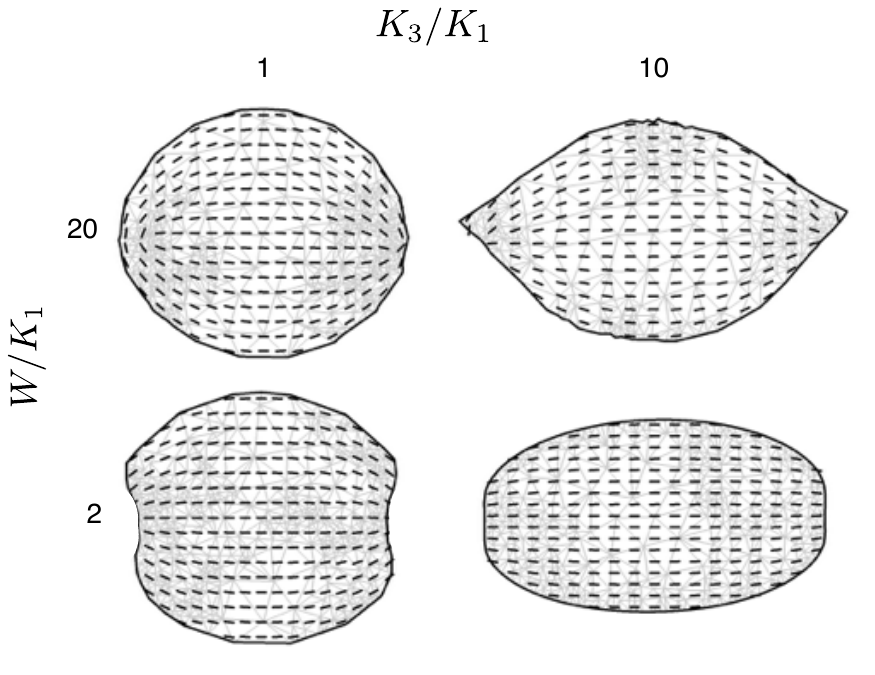}

\caption{\label{fig:Simulations-of-nematic}Simulations of nematic droplets
with weak line tension ($W/\sigma=10$). Line segments indicate the
director orientation at those points. }
\end{figure}

The effect of elastic anisotropy is shown in fig. \ref{fig:Simulations-of-nematic},
which displays four equilibrium configurations as a function of $K_{3}/K_{1}$
and $W/K_{1}$. It is clear that significant shape deformation is
only achieved with the introduction of elastic anisotropy. Stronger
elastic constants lead to a fairly homogeneous field with excluded
defects, while weak elastic constants allow the field the better follow
the anchoring condition, resulting in a pair of boojum defects. 

\begin{figure}
\includegraphics[width=1\columnwidth]{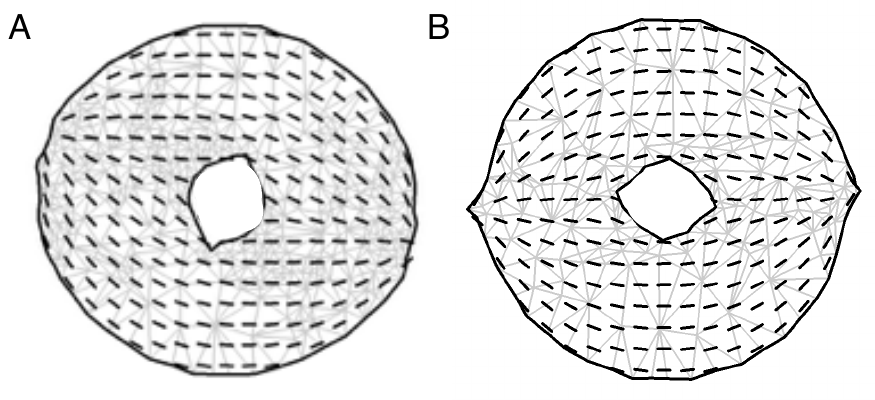}

\caption{\label{fig:Holes}Simulations of nematic droplets with weak line tension
($W/\sigma=10$) with interior holes. The holes in the droplets form
negative tactoids that align either \textbf{A} perpendicular or \textbf{B}
parallel to the orientation of the outer tactoid depending on the
ratio of $W$ to $K_{1}$. }
\end{figure}

It is straightforward to apply our technique to more complicated geometries,
such as non-simply connected domains. In figure \ref{fig:Holes},
we display two equilibrium configurations with negative tactoid holes
on the interior. These were started from an initially annular configuration
with an area constraint additionally applied to the hole. Solutions
with the interior negative tactoid aligned perpendicular to (fig.
\ref{fig:Holes}A) or parallel to (fig. \ref{fig:Holes}B) with the
exterior were found. 

We note that the array of achievable tactoid shapes fit within the
classes defined in \cite{Prinsen2003}. Additionally, the shape trends
displayed in fig. \ref{fig:Shape-of-a} closely mimic those proposed
by Lishchuk~\cite{Lishchuk2004}. Unique to our results is the ability
to predict tactoid shapes with more complex boundaries. This allows
for better agreement with experimentally produced cusped ellipsoids~\cite{Tortora2011,Kaznacheev2002},
as well as tactoids containing isotropic inclusions~\cite{Casagrande1987,Kim2013}.
Excitingly, experimental evidence of the latter tends to produce negative
tactoids that align parallel to the exterior tactoid, in agreement
with one of our classes.

\section{Conclusion\label{sec:Conclusion}}

The present work describes our initial exploration of a very challenging
class of problems, where the equilibrium configuration of a system
must be found by minimizing some functional with respect to internal
fields as well as the shape of the domain. Gradient descent methods
are effective for some subset of these problems, and form the basis
of programs like \emph{Surface Evolver}, but require management of
mesh quality during the minimization. Some of the challenges were
illustrated with a toy example of minimizing the length of a loop
at constant area, showing that mesh points become bunched during the
minimization and prevent convergence on the correct solution. 

A regularization strategy was shown to be effective in overcoming
these difficulties whereby the target functional was supplemented
with an auxiliary functional to promote equally-spaced mesh points.
The same approach was used to develop a solver for the equilibrium
configuration of a flexible capacitor, as well as that of a 2D nematic
tactoid. Exploring the space of shapes as a function of the surface
tension and anisotropic elastic constants, we found structures similar
to those experimentally observed, as well as some others predicted
in the literature. We are presently working to generalize our program
to work with higher dimensional manifolds, arbitary combinations of
fields defined upon them, a broader selection of energy functionals
and dynamics problems. 
\begin{acknowledgments}
\emph{The authors wish to thank the Tufts Graduate School of Arts
and Sciences for provision of a summer scholarship for ADB. Both TJA
and ADB are grateful for funding from a Cottrell Award from the Research
Corporation for Science Advancement. }
\end{acknowledgments}

\section*{Appendix: Conformal Mapping Solution}

The integrals in (\ref{eq:energy-1}) and (\ref{eq:areaconstraint})
must be translated into complex notation using the map $z(w)$. To
do this, we use familiar properties of complex functions, i.e that
$\frac{dz}{dw}$ plays the role of a metric, so that lengths transform
under the action of $z$ like $\left|\frac{dz}{dw}\right|$ and areas
transform like $\left|\frac{dz}{dw}\right|^{2}$. Parametrizing the
boundary of the unit circle as $e^{i\theta}$, the program in complex
notation is to minimize, 
\begin{equation}
E=\Gamma_{\sigma}\int\left|\frac{dz}{dw}\right|d\theta+\Gamma_{\epsilon}\int\left|\frac{du}{dw}\right|^{2}dw+\int(u-u_{0})\left|\frac{dz}{dw}\right|d\theta,\label{eq:complexenergy}
\end{equation}
with respect to $z(w)$ and the real function $u(w)$ and subject
to the constraint,
\begin{equation}
\frac{1}{2}\int\bar{z}\frac{dz}{dw}ie^{i\theta}d\theta=\pi,\label{eq:areaconstraintcomplex}
\end{equation}
where we used the property that the electrostatic energy is invariant
under a conformal mapping. 

The conformal map $z(w)$ and the scalar potential $u$ are expanded
as a power series in $w$,
\begin{equation}
z(w)=\sum_{n=1}^{\infty}a_{n}w^{n},\label{eq:zsoln}
\end{equation}
and
\begin{equation}
u(w)=\Im\left(\sum_{n=1}^{\infty}c_{n}w^{n}\right),\label{eq:usoln}
\end{equation}
where positive powers only have been retained to ensure regularity
at the origin $w=0$ and the imaginary part is chosen due to the symmetry
of the boundary condition. Substituting $w=re^{i\theta}$ into (\ref{eq:usoln}),
we obtain 
\begin{equation}
u(w)=\sum_{n}c_{n}r^{n}\sin(n\theta),\label{eq:usoln-1}
\end{equation}
and turn to the issue of evaluating each term in the energy functional,
as well as the area constraint, in order of difficulty.

The first term we consider is the electrostatic energy, which is after
all invariant under the mapping. Inserting the solution (\ref{eq:usoln-1})
into the second term in (\ref{eq:complexenergy}) yields,
\begin{equation}
E_{el}=\Gamma_{\epsilon}\sum_{n}\frac{2n^{2}}{2n-1}c_{n}^{2}.\label{eq:energyel}
\end{equation}
Next, we turn to the area constraint. Inserting (\ref{eq:zsoln})
into (\ref{eq:areaconstraintcomplex}) and integrating, 
\[
A=\sum_{n}n\pi a_{n}^{2}.
\]
At this point, we shall specialize the problem to only consider the
first two odd terms of $z(w)$, 
\begin{equation}
z=a_{1}w+a_{3}w^{3}\label{eq:zspecial}
\end{equation}
where the first is the identity mapping and the second represents
the first nonzero term of the deformation induced by the field. This
approximation is equivalent to assuming that the capacitor is only
weakly deformed from the circular reference shape, which is true if
the line tension dominates in the energy functional. We therefore
create a one parameter family of maps that preserves the area of the
unit disk,
\begin{equation}
z=w\sqrt{1-3\alpha^{2}}+\alpha w^{3}.\label{eq:Ourmap}
\end{equation}
The line tension term in (\ref{eq:complexenergy}) is highly nonlinear.
Inserting the restricted map (\ref{eq:Ourmap}) and evaluating using
$w=e^{i\theta}$ on the boundary, we obtain
\[
E_{lt}=\Gamma_{\sigma}\int_{-\pi}^{\pi}\sqrt{9\alpha^{2}\sin^{2}(2\theta)+\left(\sqrt{1-3\alpha^{2}}+3\alpha\cos(2\theta)\right)^{2}}d\theta,
\]
which may be evaluated in terms of the complete elliptic integral
$\text{E}(m)$,

\[
F_{lt}=2\Gamma_{\sigma}\left[G_{+}E\left(m_{+}\right)+G_{-}E\left(m_{-}\right)\right],
\]
where 
\begin{eqnarray*}
G_{\pm} & = & \sqrt{1+6\alpha\left(\alpha\pm\sqrt{1-3\alpha^{2}}\right)},\\
m_{\pm} & = & \frac{\pm12\alpha\sqrt{1-3\alpha^{2}}}{1+6\alpha\left(\alpha\pm\sqrt{1-3\alpha^{2}}\right)}.
\end{eqnarray*}
Due to the complexity of this expression, we make a series expansion
about $\alpha=0$ and retain the first nontrivial term,
\begin{equation}
E_{lt}\approx\Gamma_{\sigma}\pi\left(2+\frac{3}{2}\alpha^{2}\right).\label{eq:energylt}
\end{equation}

The anchoring term also contains $\left|\frac{dz}{dw}\right|$, and
hence is also nonlinear. We follow a similar strategy to the line
tension, expanding this in powers of $\alpha$ on the boundary $w=e^{i\theta}$
and retaining only terms up to quadratic order,
\[
\left|\frac{dz}{dw}\right|\approx1+3\alpha\cos(2\theta)-\frac{3}{4}\alpha^{2}(3\cos(4\theta)-1).
\]
Inserting this, together with the series solution for $u$, eq. (\ref{eq:usoln-1}),
into the anchoring energy term in (\ref{eq:complexenergy}) and performing
the integration yields an expression of the form,
\begin{equation}
E_{a}=\left(p+q_{i}c_{i}+r_{ij}c_{i}c_{j}\right),\label{eq:energyanchoring}
\end{equation}
where $p$, $q$ and $r$ are scalar, vector and matrix quadratic
functions of $\alpha$. $r_{ij}$ is a symmetric band-diagonal matrix
with a checkerboard structure, $r_{ij}=0$ if $\left|i-j\right|=0$.
For the sake of brevity, we do not list these elements here; they
are available in a \emph{Mathematica} notebook provided as supplementary
material. 

All terms in the free energy (\ref{eq:complexenergy}) have now been
evaluated in terms of the parameter $\alpha$ that defines the area-preserving
conformal map (\ref{eq:Ourmap}) as well as the coefficients of the
field (\ref{eq:usoln-1}). Expressions for line tension, electric
energy and anchoring are quoted in equations (\ref{eq:energylt}),
(\ref{eq:energyel}) and (\ref{eq:energyanchoring}) respectively.
It may therefore be seen that the total free energy can be put into
the form,
\begin{equation}
E=P+Q_{i}c_{i}+R_{ij}c_{i}c_{j},\label{eq:Fquadratic}
\end{equation}
where $P$, $Q$ and $R$ are quadratic functions of $\alpha$. The
scalar $P$ contains contributions from the anchoring and line tension
terms; $Q$ contains only contributions from the anchoring term; $R$
contains electrostatic (on the diagonal) and anchoring contributions.
The special structure arises because the series expansion in $z$
is low order, while the expansion in $u$ was carried out to arbitrary
order and is valid for the situation where the shape deformation is
small. A higher order expansion in $z$ will yield $P$, $Q$ and
$R$ that are also variables of higher order parameters $\{\beta,\gamma,...\}$.
Furthermore, series expansion of $\left|\frac{dz}{dw}\right|$ limits
the order of $P$, $Q$ and $R$ to being quadratic in $\alpha$;
this expansion could easily be extended further.

Minimization of eq. (\ref{eq:Fquadratic}) with respect to $\alpha$
and $c_{i}$ yields the following system of equations,
\begin{eqnarray}
\frac{\partial E}{\partial\alpha}=\frac{\partial P}{\partial\alpha}+\frac{\partial Q_{i}}{\partial\alpha}c_{i}+\frac{R_{ij}}{\partial\alpha}c_{i}c_{j} & = & 0,\nonumber \\
2R_{ij}c_{j}+Q_{i} & = & 0,\label{eq:qmin}
\end{eqnarray}
where we note that $i$ is a free index in the second equation. These
equations are efficiently solved by the nested iteration scheme,
\begin{eqnarray*}
a^{m+1} & = & \alpha^{m}-\lambda\left.\frac{\partial F}{\partial\alpha}\right|c_{i}^{m}\\
2R_{ij}(\alpha^{m})c_{j}^{m+1} & = & -Q_{i}(\alpha^{m})
\end{eqnarray*}
for sufficiently small stepsize $\lambda$ and a single initial seed
value $\alpha^{0}=0$. A \emph{Mathematica} notebook that implements
this scheme is presented as supplementary material.


\begin{thebibliography}{56}
\expandafter\ifx\csname natexlab\endcsname\relax\def\natexlab#1{#1}\fi
\expandafter\ifx\csname bibnamefont\endcsname\relax
  \def\bibnamefont#1{#1}\fi
\expandafter\ifx\csname bibfnamefont\endcsname\relax
  \def\bibfnamefont#1{#1}\fi
\expandafter\ifx\csname citenamefont\endcsname\relax
  \def\citenamefont#1{#1}\fi
\expandafter\ifx\csname url\endcsname\relax
  \def\url#1{\texttt{#1}}\fi
\expandafter\ifx\csname urlprefix\endcsname\relax\def\urlprefix{URL }\fi
\providecommand{\bibinfo}[2]{#2}
\providecommand{\eprint}[2][]{\url{#2}}

\bibitem[{\citenamefont{Barbero and Evangelista}(2005)}]{barbero2005adsorption}
\bibinfo{author}{\bibfnamefont{G.}~\bibnamefont{Barbero}} \bibnamefont{and}
  \bibinfo{author}{\bibfnamefont{L.}~\bibnamefont{Evangelista}},
  \emph{\bibinfo{title}{Adsorption Phenomena and Anchoring Energy in Nematic
  Liquid Crystals}}, Liquid Crystals Book Series (\bibinfo{publisher}{CRC
  Press}, \bibinfo{year}{2005}), ISBN \bibinfo{isbn}{9781420037456}.

\bibitem[{\citenamefont{Rasing and Musevic}(2013)}]{rasing2013surfaces}
\bibinfo{author}{\bibfnamefont{T.}~\bibnamefont{Rasing}} \bibnamefont{and}
  \bibinfo{author}{\bibfnamefont{I.}~\bibnamefont{Musevic}},
  \emph{\bibinfo{title}{Surfaces and Interfaces of Liquid Crystals}}
  (\bibinfo{publisher}{Springer Berlin Heidelberg}, \bibinfo{year}{2013}), ISBN
  \bibinfo{isbn}{9783662101575}.

\bibitem[{\citenamefont{Kim et~al.}(2002)\citenamefont{Kim, Yoneya, and
  Yokoyama}}]{Kim2002}
\bibinfo{author}{\bibfnamefont{J.-H.} \bibnamefont{Kim}},
  \bibinfo{author}{\bibfnamefont{M.}~\bibnamefont{Yoneya}}, \bibnamefont{and}
  \bibinfo{author}{\bibfnamefont{H.}~\bibnamefont{Yokoyama}},
  \bibinfo{journal}{Nature} \textbf{\bibinfo{volume}{420}},
  \bibinfo{pages}{159} (\bibinfo{year}{2002}).

\bibitem[{\citenamefont{Atherton and Adler}(2012)}]{Atherton2012}
\bibinfo{author}{\bibfnamefont{T.~J.} \bibnamefont{Atherton}} \bibnamefont{and}
  \bibinfo{author}{\bibfnamefont{J.~H.} \bibnamefont{Adler}},
  \bibinfo{journal}{Phys. Rev. E} \textbf{\bibinfo{volume}{86}},
  \bibinfo{pages}{040701} (\bibinfo{year}{2012}).

\bibitem[{\citenamefont{Kondrat et~al.}(2005)\citenamefont{Kondrat,
  Poniewierski, and Harnau}}]{Kondrat2005}
\bibinfo{author}{\bibfnamefont{S.}~\bibnamefont{Kondrat}},
  \bibinfo{author}{\bibfnamefont{a.}~\bibnamefont{Poniewierski}},
  \bibnamefont{and} \bibinfo{author}{\bibfnamefont{L.}~\bibnamefont{Harnau}},
  \bibinfo{journal}{Liq. Cryst.} \textbf{\bibinfo{volume}{32}},
  \bibinfo{pages}{95} (\bibinfo{year}{2005}).

\bibitem[{\citenamefont{Anquetil-Deck et~al.}(2013)\citenamefont{Anquetil-Deck,
  Cleaver, Bramble, and Atherton}}]{Anquetil-Deck2013}
\bibinfo{author}{\bibfnamefont{C.}~\bibnamefont{Anquetil-Deck}},
  \bibinfo{author}{\bibfnamefont{D.~J.} \bibnamefont{Cleaver}},
  \bibinfo{author}{\bibfnamefont{J.~P.} \bibnamefont{Bramble}},
  \bibnamefont{and} \bibinfo{author}{\bibfnamefont{T.~J.}
  \bibnamefont{Atherton}}, \bibinfo{journal}{Phys. Rev. E}
  \textbf{\bibinfo{volume}{88}}, \bibinfo{pages}{012501}
  (\bibinfo{year}{2013}).

\bibitem[{\citenamefont{Lagerwall and Scalia}(2012)}]{lagerwall2012new}
\bibinfo{author}{\bibfnamefont{J.~P.} \bibnamefont{Lagerwall}}
  \bibnamefont{and} \bibinfo{author}{\bibfnamefont{G.}~\bibnamefont{Scalia}},
  \bibinfo{journal}{Current Applied Physics} \textbf{\bibinfo{volume}{12}},
  \bibinfo{pages}{1387} (\bibinfo{year}{2012}).

\bibitem[{\citenamefont{MacKintosh and
  Lubensky}(1991)}]{MacKintoshLubensky1991}
\bibinfo{author}{\bibfnamefont{F.~C.} \bibnamefont{MacKintosh}}
  \bibnamefont{and} \bibinfo{author}{\bibfnamefont{T.~C.}
  \bibnamefont{Lubensky}}, \bibinfo{journal}{Phys. Rev. Lett.}
  \textbf{\bibinfo{volume}{{\bf 67}}}, \bibinfo{pages}{1169}
  (\bibinfo{year}{1991}).

\bibitem[{\citenamefont{Fern{\'a}ndez-Nieves
  et~al.}(2007)\citenamefont{Fern{\'a}ndez-Nieves, Vitelli, Utada, Link,
  M{\'a}rquez, Nelson, and Weitz}}]{fernandez2007novel}
\bibinfo{author}{\bibfnamefont{A.}~\bibnamefont{Fern{\'a}ndez-Nieves}},
  \bibinfo{author}{\bibfnamefont{V.}~\bibnamefont{Vitelli}},
  \bibinfo{author}{\bibfnamefont{A.~S.} \bibnamefont{Utada}},
  \bibinfo{author}{\bibfnamefont{D.~R.} \bibnamefont{Link}},
  \bibinfo{author}{\bibfnamefont{M.}~\bibnamefont{M{\'a}rquez}},
  \bibinfo{author}{\bibfnamefont{D.~R.} \bibnamefont{Nelson}},
  \bibnamefont{and} \bibinfo{author}{\bibfnamefont{D.~A.} \bibnamefont{Weitz}},
  \bibinfo{journal}{Physical review letters} \textbf{\bibinfo{volume}{99}},
  \bibinfo{pages}{157801} (\bibinfo{year}{2007}).

\bibitem[{\citenamefont{Whitmer et~al.}(2013)\citenamefont{Whitmer, Wang,
  Mondiot, Miller, Abbott, and de~Pablo}}]{PhysRevLett.111.227801}
\bibinfo{author}{\bibfnamefont{J.~K.} \bibnamefont{Whitmer}},
  \bibinfo{author}{\bibfnamefont{X.}~\bibnamefont{Wang}},
  \bibinfo{author}{\bibfnamefont{F.}~\bibnamefont{Mondiot}},
  \bibinfo{author}{\bibfnamefont{D.~S.} \bibnamefont{Miller}},
  \bibinfo{author}{\bibfnamefont{N.~L.} \bibnamefont{Abbott}},
  \bibnamefont{and} \bibinfo{author}{\bibfnamefont{J.~J.}
  \bibnamefont{de~Pablo}}, \bibinfo{journal}{Phys. Rev. Lett.}
  \textbf{\bibinfo{volume}{111}}, \bibinfo{pages}{227801}
  (\bibinfo{year}{2013}).

\bibitem[{\citenamefont{Zocher}(1925)}]{Zocher1925}
\bibinfo{author}{\bibfnamefont{V.~H.} \bibnamefont{Zocher}},
  \bibinfo{journal}{Zeitschrift f\"{u}r Anorg. und Allg. Chemie}
  \textbf{\bibinfo{volume}{147}}, \bibinfo{pages}{91} (\bibinfo{year}{1925}),
  ISSN \bibinfo{issn}{1521-3749}.

\bibitem[{\citenamefont{Zocher and Jacobsohn}(1929)}]{Zocher1929}
\bibinfo{author}{\bibfnamefont{V.~H.} \bibnamefont{Zocher}} \bibnamefont{and}
  \bibinfo{author}{\bibfnamefont{K.}~\bibnamefont{Jacobsohn}},
  \bibinfo{journal}{Kolloid Beih.} \textbf{\bibinfo{volume}{28}},
  \bibinfo{pages}{167} (\bibinfo{year}{1929}).

\bibitem[{\citenamefont{Bernal and Fankuchen}(1941)}]{Bernal1941}
\bibinfo{author}{\bibfnamefont{J.~D.} \bibnamefont{Bernal}} \bibnamefont{and}
  \bibinfo{author}{\bibfnamefont{I.}~\bibnamefont{Fankuchen}},
  \bibinfo{journal}{J. Gen. Physiol.} \textbf{\bibinfo{volume}{25}},
  \bibinfo{pages}{111} (\bibinfo{year}{1941}), ISSN \bibinfo{issn}{0022-1295}.

\bibitem[{\citenamefont{Sonin}(1998)}]{Sonin1998}
\bibinfo{author}{\bibfnamefont{A.~S.} \bibnamefont{Sonin}},
  \bibinfo{journal}{J. Mater. Chem.} \textbf{\bibinfo{volume}{8}},
  \bibinfo{pages}{2557} (\bibinfo{year}{1998}), ISSN \bibinfo{issn}{09599428}.

\bibitem[{\citenamefont{Tang et~al.}(2005)\citenamefont{Tang, Kang, and
  Jia}}]{Tang2005}
\bibinfo{author}{\bibfnamefont{J.~X.} \bibnamefont{Tang}},
  \bibinfo{author}{\bibfnamefont{H.}~\bibnamefont{Kang}}, \bibnamefont{and}
  \bibinfo{author}{\bibfnamefont{J.}~\bibnamefont{Jia}},
  \bibinfo{journal}{Langmuir} \textbf{\bibinfo{volume}{21}},
  \bibinfo{pages}{2789} (\bibinfo{year}{2005}), ISSN \bibinfo{issn}{07437463}.

\bibitem[{\citenamefont{Casagrande et~al.}(1987)\citenamefont{Casagrande,
  Fabre, Guedeau, and Veyssie}}]{Casagrande1987}
\bibinfo{author}{\bibfnamefont{C.}~\bibnamefont{Casagrande}},
  \bibinfo{author}{\bibfnamefont{P.}~\bibnamefont{Fabre}},
  \bibinfo{author}{\bibfnamefont{M.~A.} \bibnamefont{Guedeau}},
  \bibnamefont{and} \bibinfo{author}{\bibfnamefont{M.}~\bibnamefont{Veyssie}},
  \bibinfo{journal}{Europhys. Lett.} \textbf{\bibinfo{volume}{3}},
  \bibinfo{pages}{73} (\bibinfo{year}{1987}), ISSN \bibinfo{issn}{0295-5075}.

\bibitem[{\citenamefont{Kim et~al.}(2013)\citenamefont{Kim, Shiyanovskii, and
  Lavrentovich}}]{Kim2013}
\bibinfo{author}{\bibfnamefont{Y.-K.} \bibnamefont{Kim}},
  \bibinfo{author}{\bibfnamefont{S.~V.} \bibnamefont{Shiyanovskii}},
  \bibnamefont{and} \bibinfo{author}{\bibfnamefont{O.~D.}
  \bibnamefont{Lavrentovich}}, \bibinfo{journal}{J. Phys. Condens. Matter}
  \textbf{\bibinfo{volume}{25}}, \bibinfo{pages}{404202}
  (\bibinfo{year}{2013}), ISSN \bibinfo{issn}{1361-648X}, \eprint{1303.6239}.

\bibitem[{\citenamefont{Verhoeff et~al.}(2011)\citenamefont{Verhoeff, Bakelaar,
  Otten, {Van Der Schoot}, and Lekkerkerker}}]{Verhoeff2011}
\bibinfo{author}{\bibfnamefont{A.~A.} \bibnamefont{Verhoeff}},
  \bibinfo{author}{\bibfnamefont{I.~A.} \bibnamefont{Bakelaar}},
  \bibinfo{author}{\bibfnamefont{R.~H.~J.} \bibnamefont{Otten}},
  \bibinfo{author}{\bibfnamefont{P.}~\bibnamefont{{Van Der Schoot}}},
  \bibnamefont{and} \bibinfo{author}{\bibfnamefont{H.~N.~W.}
  \bibnamefont{Lekkerkerker}}, \bibinfo{journal}{Langmuir}
  \textbf{\bibinfo{volume}{27}}, \bibinfo{pages}{116} (\bibinfo{year}{2011}),
  ISSN \bibinfo{issn}{07437463}.

\bibitem[{\citenamefont{Peng and Lavrentovich}(2015)}]{Peng2015}
\bibinfo{author}{\bibfnamefont{C.}~\bibnamefont{Peng}} \bibnamefont{and}
  \bibinfo{author}{\bibfnamefont{O.~D.} \bibnamefont{Lavrentovich}},
  \bibinfo{journal}{Soft Matter} \textbf{\bibinfo{volume}{11}},
  \bibinfo{pages}{10} (\bibinfo{year}{2015}), ISSN \bibinfo{issn}{1744-683X},
  \eprint{1507.07499}.

\bibitem[{\citenamefont{Mushenheim et~al.}(2014)\citenamefont{Mushenheim,
  Trivedi, Weibel, and Abbott}}]{Mushenheim2014}
\bibinfo{author}{\bibfnamefont{P.~C.} \bibnamefont{Mushenheim}},
  \bibinfo{author}{\bibfnamefont{R.~R.} \bibnamefont{Trivedi}},
  \bibinfo{author}{\bibfnamefont{D.~B.} \bibnamefont{Weibel}},
  \bibnamefont{and} \bibinfo{author}{\bibfnamefont{N.~L.}
  \bibnamefont{Abbott}}, \bibinfo{journal}{Biophys. J.}
  \textbf{\bibinfo{volume}{107}}, \bibinfo{pages}{255} (\bibinfo{year}{2014}),
  ISSN \bibinfo{issn}{15420086}.

\bibitem[{\citenamefont{Zhou et~al.}(2014)\citenamefont{Zhou, Sokolov,
  Lavrentovich, and Aranson}}]{Zhou2014}
\bibinfo{author}{\bibfnamefont{S.}~\bibnamefont{Zhou}},
  \bibinfo{author}{\bibfnamefont{A.}~\bibnamefont{Sokolov}},
  \bibinfo{author}{\bibfnamefont{O.~D.} \bibnamefont{Lavrentovich}},
  \bibnamefont{and} \bibinfo{author}{\bibfnamefont{I.~S.}
  \bibnamefont{Aranson}}, \bibinfo{journal}{Proc. Natl. Acad. Sci. U. S. A.}
  \textbf{\bibinfo{volume}{111}}, \bibinfo{pages}{1265} (\bibinfo{year}{2014}),
  ISSN \bibinfo{issn}{1091-6490}, \eprint{1312.5359}.

\bibitem[{\citenamefont{Dogic and Fraden}(2001)}]{Dogic2001}
\bibinfo{author}{\bibfnamefont{Z.}~\bibnamefont{Dogic}} \bibnamefont{and}
  \bibinfo{author}{\bibfnamefont{S.}~\bibnamefont{Fraden}},
  \bibinfo{journal}{Philos. Trans. R. Soc. A Math. Phys. Eng. Sci.}
  \textbf{\bibinfo{volume}{359}}, \bibinfo{pages}{997} (\bibinfo{year}{2001}),
  ISSN \bibinfo{issn}{1364-503X}.

\bibitem[{\citenamefont{Chandrasekhar}(1966)}]{Chandrasekhar1966}
\bibinfo{author}{\bibfnamefont{S.}~\bibnamefont{Chandrasekhar}},
  \bibinfo{journal}{Mol. Cryst. Liq. Cryst.} \textbf{\bibinfo{volume}{2}},
  \bibinfo{pages}{71} (\bibinfo{year}{1966}), ISSN \bibinfo{issn}{0369-1152}.

\bibitem[{\citenamefont{Kaznacheev et~al.}(2003)\citenamefont{Kaznacheev,
  Bogdanov, and Sonin}}]{Kaznacheev2003}
\bibinfo{author}{\bibfnamefont{A.~V.} \bibnamefont{Kaznacheev}},
  \bibinfo{author}{\bibfnamefont{M.~M.} \bibnamefont{Bogdanov}},
  \bibnamefont{and} \bibinfo{author}{\bibfnamefont{A.~S.} \bibnamefont{Sonin}},
  \bibinfo{journal}{J. Exp. Theor. Phys.} \textbf{\bibinfo{volume}{97}},
  \bibinfo{pages}{1159} (\bibinfo{year}{2003}), ISSN \bibinfo{issn}{1063-7761}.

\bibitem[{\citenamefont{Kaznacheev et~al.}(2002)\citenamefont{Kaznacheev,
  Bogdanov, and Taraskin}}]{Kaznacheev2002}
\bibinfo{author}{\bibfnamefont{A.~V.} \bibnamefont{Kaznacheev}},
  \bibinfo{author}{\bibfnamefont{M.~M.} \bibnamefont{Bogdanov}},
  \bibnamefont{and} \bibinfo{author}{\bibfnamefont{S.~A.}
  \bibnamefont{Taraskin}}, \bibinfo{journal}{J. Exp. Theor. Phys.}
  \textbf{\bibinfo{volume}{95}}, \bibinfo{pages}{57} (\bibinfo{year}{2002}).

\bibitem[{\citenamefont{Prinsen and van~der Schoot}(2003)}]{Prinsen2003}
\bibinfo{author}{\bibfnamefont{P.}~\bibnamefont{Prinsen}} \bibnamefont{and}
  \bibinfo{author}{\bibfnamefont{P.}~\bibnamefont{van~der Schoot}},
  \bibinfo{journal}{Phys. Rev. E. Stat. Nonlin. Soft Matter Phys.}
  \textbf{\bibinfo{volume}{68}}, \bibinfo{pages}{021701}
  (\bibinfo{year}{2003}), ISSN \bibinfo{issn}{1063-651X}.

\bibitem[{\citenamefont{Prinsen and {Van Der Schoot}}(2004)}]{Prinsen2004a}
\bibinfo{author}{\bibfnamefont{P.}~\bibnamefont{Prinsen}} \bibnamefont{and}
  \bibinfo{author}{\bibfnamefont{P.}~\bibnamefont{{Van Der Schoot}}},
  \bibinfo{journal}{Eur. Phys. J. E} \textbf{\bibinfo{volume}{13}},
  \bibinfo{pages}{35} (\bibinfo{year}{2004}), ISSN \bibinfo{issn}{12928941}.

\bibitem[{\citenamefont{Bates}(2003)}]{Bates2003}
\bibinfo{author}{\bibfnamefont{M.~A.} \bibnamefont{Bates}},
  \bibinfo{journal}{Chem. Phys. Lett.} \textbf{\bibinfo{volume}{368}},
  \bibinfo{pages}{87} (\bibinfo{year}{2003}), ISSN \bibinfo{issn}{00092614}.

\bibitem[{\citenamefont{Vanzo et~al.}(2012)\citenamefont{Vanzo, Ricci, Berardi,
  and Zannoni}}]{Vanzo2012}
\bibinfo{author}{\bibfnamefont{D.}~\bibnamefont{Vanzo}},
  \bibinfo{author}{\bibfnamefont{M.}~\bibnamefont{Ricci}},
  \bibinfo{author}{\bibfnamefont{R.}~\bibnamefont{Berardi}}, \bibnamefont{and}
  \bibinfo{author}{\bibfnamefont{C.}~\bibnamefont{Zannoni}},
  \bibinfo{journal}{Soft Matter} \textbf{\bibinfo{volume}{8}},
  \bibinfo{pages}{11790} (\bibinfo{year}{2012}), ISSN
  \bibinfo{issn}{1744-683X}.

\bibitem[{\citenamefont{Warner and Terentjev}(2003{\natexlab{a}})}]{Warnerbook}
\bibinfo{author}{\bibfnamefont{M.}~\bibnamefont{Warner}} \bibnamefont{and}
  \bibinfo{author}{\bibfnamefont{E.}~\bibnamefont{Terentjev}},
  \emph{\bibinfo{title}{Liquid Crystal Elastomers}} (\bibinfo{publisher}{Oxford
  Science Publications}, \bibinfo{year}{2003}{\natexlab{a}}).

\bibitem[{\citenamefont{De~Gennes}(1975)}]{DeGennes}
\bibinfo{author}{\bibfnamefont{P.~G.} \bibnamefont{De~Gennes}},
  \bibinfo{journal}{C. R. Acad. Sci. B} \textbf{\bibinfo{volume}{281}},
  \bibinfo{pages}{101} (\bibinfo{year}{1975}).

\bibitem[{\citenamefont{Kundler and Finkelmann}(1995)}]{kundler_finkelmann}
\bibinfo{author}{\bibfnamefont{I.}~\bibnamefont{Kundler}} \bibnamefont{and}
  \bibinfo{author}{\bibfnamefont{H.}~\bibnamefont{Finkelmann}},
  \bibinfo{journal}{Macromol. Rapid Commun.} \textbf{\bibinfo{volume}{16}},
  \bibinfo{pages}{679} (\bibinfo{year}{1995}).

\bibitem[{\citenamefont{Finkelmann et~al.}(2001)\citenamefont{Finkelmann,
  Nishikawa, Pereira, and Warner}}]{Finkelmann2001}
\bibinfo{author}{\bibfnamefont{H.}~\bibnamefont{Finkelmann}},
  \bibinfo{author}{\bibfnamefont{E.}~\bibnamefont{Nishikawa}},
  \bibinfo{author}{\bibfnamefont{G.}~\bibnamefont{Pereira}}, \bibnamefont{and}
  \bibinfo{author}{\bibfnamefont{M.}~\bibnamefont{Warner}},
  \bibinfo{journal}{Physical Review Letters} \textbf{\bibinfo{volume}{87}},
  \bibinfo{pages}{015501} (\bibinfo{year}{2001}), ISSN
  \bibinfo{issn}{0031-9007}.

\bibitem[{\citenamefont{Conti et~al.}(2002)\citenamefont{Conti, DeSimone, and
  Dolzmann}}]{Conti2002}
\bibinfo{author}{\bibfnamefont{S.}~\bibnamefont{Conti}},
  \bibinfo{author}{\bibfnamefont{A.}~\bibnamefont{DeSimone}}, \bibnamefont{and}
  \bibinfo{author}{\bibfnamefont{G.}~\bibnamefont{Dolzmann}},
  \bibinfo{journal}{Physical Review E} \textbf{\bibinfo{volume}{66}},
  \bibinfo{pages}{061710} (\bibinfo{year}{2002}), ISSN
  \bibinfo{issn}{1063-651X}.

\bibitem[{\citenamefont{Mbanga et~al.}(2010{\natexlab{a}})\citenamefont{Mbanga,
  Ye, Selinger, and Selinger}}]{Mbanga_Selinger_pre}
\bibinfo{author}{\bibfnamefont{B.~L.} \bibnamefont{Mbanga}},
  \bibinfo{author}{\bibfnamefont{F.}~\bibnamefont{Ye}},
  \bibinfo{author}{\bibfnamefont{J.~V.} \bibnamefont{Selinger}},
  \bibnamefont{and} \bibinfo{author}{\bibfnamefont{R.~L.~B.}
  \bibnamefont{Selinger}}, \bibinfo{journal}{Phys. Rev. E}
  \textbf{\bibinfo{volume}{82}}, \bibinfo{pages}{051701}
  (\bibinfo{year}{2010}{\natexlab{a}}).

\bibitem[{\citenamefont{Chang et~al.}(1997)\citenamefont{Chang, Chien, and
  Meyer}}]{Chang1997}
\bibinfo{author}{\bibfnamefont{C.-C.} \bibnamefont{Chang}},
  \bibinfo{author}{\bibfnamefont{L.-C.} \bibnamefont{Chien}}, \bibnamefont{and}
  \bibinfo{author}{\bibfnamefont{R.}~\bibnamefont{Meyer}},
  \bibinfo{journal}{Physical Review E} \textbf{\bibinfo{volume}{56}},
  \bibinfo{pages}{595} (\bibinfo{year}{1997}), ISSN \bibinfo{issn}{1063-651X}.

\bibitem[{\citenamefont{Burridge et~al.}(2006)\citenamefont{Burridge, Mao, and
  Warner}}]{Burridge2006}
\bibinfo{author}{\bibfnamefont{D.}~\bibnamefont{Burridge}},
  \bibinfo{author}{\bibfnamefont{Y.}~\bibnamefont{Mao}}, \bibnamefont{and}
  \bibinfo{author}{\bibfnamefont{M.}~\bibnamefont{Warner}},
  \bibinfo{journal}{Physical Review E} \textbf{\bibinfo{volume}{74}},
  \bibinfo{pages}{021708} (\bibinfo{year}{2006}), ISSN
  \bibinfo{issn}{1539-3755}.

\bibitem[{\citenamefont{Menzel and Brand}(2007)}]{Menzel2007}
\bibinfo{author}{\bibfnamefont{A.}~\bibnamefont{Menzel}} \bibnamefont{and}
  \bibinfo{author}{\bibfnamefont{H.}~\bibnamefont{Brand}},
  \bibinfo{journal}{Physical Review E} \textbf{\bibinfo{volume}{75}},
  \bibinfo{pages}{011707} (\bibinfo{year}{2007}), ISSN
  \bibinfo{issn}{1539-3755}.

\bibitem[{\citenamefont{Ye et~al.}(2007)\citenamefont{Ye, Mukhopadhyay,
  Stenull, and Lubensky}}]{Ye2007}
\bibinfo{author}{\bibfnamefont{F.}~\bibnamefont{Ye}},
  \bibinfo{author}{\bibfnamefont{R.}~\bibnamefont{Mukhopadhyay}},
  \bibinfo{author}{\bibfnamefont{O.}~\bibnamefont{Stenull}}, \bibnamefont{and}
  \bibinfo{author}{\bibfnamefont{T.}~\bibnamefont{Lubensky}},
  \bibinfo{journal}{Physical Review Letters} \textbf{\bibinfo{volume}{98}},
  \bibinfo{pages}{147801} (\bibinfo{year}{2007}), ISSN
  \bibinfo{issn}{0031-9007}.

\bibitem[{\citenamefont{Urayama et~al.}(2003)\citenamefont{Urayama, Okuno, and
  Kohjiya}}]{Urayama2003}
\bibinfo{author}{\bibfnamefont{K.}~\bibnamefont{Urayama}},
  \bibinfo{author}{\bibfnamefont{Y.}~\bibnamefont{Okuno}}, \bibnamefont{and}
  \bibinfo{author}{\bibfnamefont{S.}~\bibnamefont{Kohjiya}},
  \bibinfo{journal}{Macromolecules} \textbf{\bibinfo{volume}{36}},
  \bibinfo{pages}{6229} (\bibinfo{year}{2003}), ISSN \bibinfo{issn}{0024-9297}.

\bibitem[{\citenamefont{Urayama et~al.}(2006)\citenamefont{Urayama, Mashita,
  Arai, and Takigawa}}]{Urayama2006}
\bibinfo{author}{\bibfnamefont{K.}~\bibnamefont{Urayama}},
  \bibinfo{author}{\bibfnamefont{R.}~\bibnamefont{Mashita}},
  \bibinfo{author}{\bibfnamefont{Y.~O.} \bibnamefont{Arai}}, \bibnamefont{and}
  \bibinfo{author}{\bibfnamefont{T.}~\bibnamefont{Takigawa}},
  \bibinfo{journal}{Macromolecules} \textbf{\bibinfo{volume}{39}},
  \bibinfo{pages}{8511} (\bibinfo{year}{2006}), ISSN \bibinfo{issn}{0024-9297}.

\bibitem[{\citenamefont{Yu et~al.}(2003)\citenamefont{Yu, Nakano, and
  Ikeda}}]{Yu2003}
\bibinfo{author}{\bibfnamefont{Y.}~\bibnamefont{Yu}},
  \bibinfo{author}{\bibfnamefont{M.}~\bibnamefont{Nakano}}, \bibnamefont{and}
  \bibinfo{author}{\bibfnamefont{T.}~\bibnamefont{Ikeda}},
  \bibinfo{journal}{Nature} \textbf{\bibinfo{volume}{425}},
  \bibinfo{pages}{145} (\bibinfo{year}{2003}), ISSN \bibinfo{issn}{1476-4687}.

\bibitem[{\citenamefont{Li et~al.}(2003)\citenamefont{Li, Keller, Li, Wang, and
  Brunet}}]{azo_Keller_2003}
\bibinfo{author}{\bibfnamefont{M.-H.} \bibnamefont{Li}},
  \bibinfo{author}{\bibfnamefont{P.}~\bibnamefont{Keller}},
  \bibinfo{author}{\bibfnamefont{B.}~\bibnamefont{Li}},
  \bibinfo{author}{\bibfnamefont{X.}~\bibnamefont{Wang}}, \bibnamefont{and}
  \bibinfo{author}{\bibfnamefont{M.}~\bibnamefont{Brunet}},
  \bibinfo{journal}{Advanced Materials} \textbf{\bibinfo{volume}{15}},
  \bibinfo{pages}{569} (\bibinfo{year}{2003}), ISSN \bibinfo{issn}{1521-4095}.

\bibitem[{\citenamefont{Camacho-Lopez et~al.}(2004)\citenamefont{Camacho-Lopez,
  Finkelmann, Palffy-Muhoray, and Shelley}}]{Camacho-Lopez2004}
\bibinfo{author}{\bibfnamefont{M.}~\bibnamefont{Camacho-Lopez}},
  \bibinfo{author}{\bibfnamefont{H.}~\bibnamefont{Finkelmann}},
  \bibinfo{author}{\bibfnamefont{P.}~\bibnamefont{Palffy-Muhoray}},
  \bibnamefont{and} \bibinfo{author}{\bibfnamefont{M.~J.}
  \bibnamefont{Shelley}}, \bibinfo{journal}{Nature materials}
  \textbf{\bibinfo{volume}{3}}, \bibinfo{pages}{307} (\bibinfo{year}{2004}),
  ISSN \bibinfo{issn}{1476-1122}.

\bibitem[{\citenamefont{He}(2007)}]{He2007}
\bibinfo{author}{\bibfnamefont{L.}~\bibnamefont{He}},
  \bibinfo{journal}{Physical Review E} \textbf{\bibinfo{volume}{75}},
  \bibinfo{pages}{041702} (\bibinfo{year}{2007}), ISSN
  \bibinfo{issn}{1539-3755}.

\bibitem[{\citenamefont{Sawa et~al.}(2010)\citenamefont{Sawa, Urayama,
  Takigawa, DeSimone, and Teresi}}]{Sawa2010}
\bibinfo{author}{\bibfnamefont{Y.}~\bibnamefont{Sawa}},
  \bibinfo{author}{\bibfnamefont{K.}~\bibnamefont{Urayama}},
  \bibinfo{author}{\bibfnamefont{T.}~\bibnamefont{Takigawa}},
  \bibinfo{author}{\bibfnamefont{A.}~\bibnamefont{DeSimone}}, \bibnamefont{and}
  \bibinfo{author}{\bibfnamefont{L.}~\bibnamefont{Teresi}},
  \bibinfo{journal}{Macromolecules} \textbf{\bibinfo{volume}{43}},
  \bibinfo{pages}{4362} (\bibinfo{year}{2010}), ISSN \bibinfo{issn}{0024-9297}.

\bibitem[{\citenamefont{Mahimwalla et~al.}(2012)\citenamefont{Mahimwalla,
  Yager, Mamiya, Shishido, Priimagi, and Barrett}}]{azo_review_Barrett}
\bibinfo{author}{\bibfnamefont{Z.}~\bibnamefont{Mahimwalla}},
  \bibinfo{author}{\bibfnamefont{K.}~\bibnamefont{Yager}},
  \bibinfo{author}{\bibfnamefont{J.-i.} \bibnamefont{Mamiya}},
  \bibinfo{author}{\bibfnamefont{A.}~\bibnamefont{Shishido}},
  \bibinfo{author}{\bibfnamefont{A.}~\bibnamefont{Priimagi}}, \bibnamefont{and}
  \bibinfo{author}{\bibfnamefont{C.}~\bibnamefont{Barrett}},
  \bibinfo{journal}{Polymer Bulletin} \textbf{\bibinfo{volume}{69}},
  \bibinfo{pages}{967} (\bibinfo{year}{2012}), ISSN \bibinfo{issn}{0170-0839}.

\bibitem[{\citenamefont{Yu et~al.}(2002)\citenamefont{Yu, Nakano, and
  Ikeda}}]{Ikeda}
\bibinfo{author}{\bibfnamefont{Y.}~\bibnamefont{Yu}},
  \bibinfo{author}{\bibfnamefont{M.}~\bibnamefont{Nakano}}, \bibnamefont{and}
  \bibinfo{author}{\bibfnamefont{T.}~\bibnamefont{Ikeda}},
  \bibinfo{journal}{Nature} \textbf{\bibinfo{volume}{425}},
  \bibinfo{pages}{145} (\bibinfo{year}{2002}).

\bibitem[{\citenamefont{Warner and
  Terentjev}(2003{\natexlab{b}})}]{MASY:MASY200351008}
\bibinfo{author}{\bibfnamefont{M.}~\bibnamefont{Warner}} \bibnamefont{and}
  \bibinfo{author}{\bibfnamefont{E.}~\bibnamefont{Terentjev}},
  \bibinfo{journal}{Macromolecular Symposia} \textbf{\bibinfo{volume}{200}},
  \bibinfo{pages}{81} (\bibinfo{year}{2003}{\natexlab{b}}), ISSN
  \bibinfo{issn}{1521-3900}.

\bibitem[{\citenamefont{Palfffy-Muhoray}(2009)}]{Palffy-Muhoray2009}
\bibinfo{author}{\bibfnamefont{P.}~\bibnamefont{Palfffy-Muhoray}},
  \bibinfo{journal}{Nat. Mater.} \textbf{\bibinfo{volume}{\textbf{8}}},
  \bibinfo{pages}{614} (\bibinfo{year}{2009}).

\bibitem[{\citenamefont{Xie and Zhang}(2005)}]{smart_materials}
\bibinfo{author}{\bibfnamefont{P.}~\bibnamefont{Xie}} \bibnamefont{and}
  \bibinfo{author}{\bibfnamefont{R.}~\bibnamefont{Zhang}}, \bibinfo{journal}{J.
  Mater. Chem.} \textbf{\bibinfo{volume}{15}}, \bibinfo{pages}{2529}
  (\bibinfo{year}{2005}).

\bibitem[{\citenamefont{Mbanga et~al.}(2010{\natexlab{b}})\citenamefont{Mbanga,
  Ye, Selinger, and Selinger}}]{mbanga2010modeling}
\bibinfo{author}{\bibfnamefont{B.~L.} \bibnamefont{Mbanga}},
  \bibinfo{author}{\bibfnamefont{F.}~\bibnamefont{Ye}},
  \bibinfo{author}{\bibfnamefont{J.~V.} \bibnamefont{Selinger}},
  \bibnamefont{and} \bibinfo{author}{\bibfnamefont{R.~L.}
  \bibnamefont{Selinger}}, \bibinfo{journal}{Physical Review E}
  \textbf{\bibinfo{volume}{82}}, \bibinfo{pages}{051701}
  (\bibinfo{year}{2010}{\natexlab{b}}).

\bibitem[{\citenamefont{de~Haan et~al.}(2014)\citenamefont{de~Haan,
  Gimenez-Pinto, Konya, Nguyen, Verjans, Sánchez-Somolinos, Selinger,
  Selinger, Broer, and Schenning}}]{ADFM:ADFM201302568}
\bibinfo{author}{\bibfnamefont{L.~T.} \bibnamefont{de~Haan}},
  \bibinfo{author}{\bibfnamefont{V.}~\bibnamefont{Gimenez-Pinto}},
  \bibinfo{author}{\bibfnamefont{A.}~\bibnamefont{Konya}},
  \bibinfo{author}{\bibfnamefont{T.-S.} \bibnamefont{Nguyen}},
  \bibinfo{author}{\bibfnamefont{J.~M.~N.} \bibnamefont{Verjans}},
  \bibinfo{author}{\bibfnamefont{C.}~\bibnamefont{Sánchez-Somolinos}},
  \bibinfo{author}{\bibfnamefont{J.~V.} \bibnamefont{Selinger}},
  \bibinfo{author}{\bibfnamefont{R.~L.~B.} \bibnamefont{Selinger}},
  \bibinfo{author}{\bibfnamefont{D.~J.} \bibnamefont{Broer}}, \bibnamefont{and}
  \bibinfo{author}{\bibfnamefont{A.~P. H.~J.} \bibnamefont{Schenning}},
  \bibinfo{journal}{Advanced Functional Materials}
  \textbf{\bibinfo{volume}{24}}, \bibinfo{pages}{1251} (\bibinfo{year}{2014}),
  ISSN \bibinfo{issn}{1616-3028}.

\bibitem[{\citenamefont{Brakke}(1992)}]{brakke1992surface}
\bibinfo{author}{\bibfnamefont{K.~A.} \bibnamefont{Brakke}},
  \bibinfo{journal}{Experimental mathematics} \textbf{\bibinfo{volume}{1}},
  \bibinfo{pages}{141} (\bibinfo{year}{1992}).

\bibitem[{\citenamefont{Lishchuk and Care}(2004)}]{Lishchuk2004}
\bibinfo{author}{\bibfnamefont{S.~V.} \bibnamefont{Lishchuk}} \bibnamefont{and}
  \bibinfo{author}{\bibfnamefont{C.~M.} \bibnamefont{Care}},
  \bibinfo{journal}{Phys. Rev. E - Stat. Nonlinear, Soft Matter Phys.}
  \textbf{\bibinfo{volume}{70}}, \bibinfo{pages}{1} (\bibinfo{year}{2004}),
  ISSN \bibinfo{issn}{15393755}.

\bibitem[{\citenamefont{Tortora and Lavrentovich}(2011)}]{Tortora2011}
\bibinfo{author}{\bibfnamefont{L.}~\bibnamefont{Tortora}} \bibnamefont{and}
  \bibinfo{author}{\bibfnamefont{O.~D.} \bibnamefont{Lavrentovich}},
  \bibinfo{journal}{Proc. Natl. Acad. Sci. U. S. A.}
  \textbf{\bibinfo{volume}{108}}, \bibinfo{pages}{5163} (\bibinfo{year}{2011}),
  ISSN \bibinfo{issn}{0027-8424}.

\end{thebibliography}
\end{document}